\documentclass[twocolumn,showpacs,aps,prl]{revtex4-1}

\usepackage{amsmath,amsfonts,amssymb}
\usepackage{wrapfig}
\usepackage{graphicx}
\usepackage{bbm}
\usepackage{color}

\begin{document}

\title{Revealing Tripartite Quantum Discord with Tripartite Information Diagram}

\author{Wei-Ting Lee}
\author{Che-Ming Li}
\email{cmli@mail.ncku.edu.tw}
\affiliation{Department of Engineering Science, National Cheng Kung University, Tainan 70101, Taiwan}

\begin{abstract}
A new measure based on the tripartite information diagram is proposed for identifying quantum discord in tripartite systems. The proposed measure generalizes the mutual information underlying discord from bipartite to tripartite systems, and utilizes both one-particle and two-particle projective measurements to reveal the characteristics of the tripartite quantum discord. The~feasibility of the proposed measure is demonstrated by evaluating the tripartite quantum discord for systems with states close to Greenberger--Horne--Zeilinger, W, and biseparable states. In~addition, the~connections between tripartite quantum discord and two other quantum correlations---namely~genuine tripartite entanglement and genuine tripartite Einstein--Podolsky--Rosen steering---are briefly discussed. The~present study considers the case of quantum discord in tripartite systems. However, the proposed framework can be readily extended to general \emph{N}-partite~systems.
\end{abstract}

\maketitle

\section{Introduction}

Quantum correlation plays an important role in investigating quantum physics and its associated behaviors and applications, such as quantum critical phenomena \cite{critical}, quantum evolution under decoherence \cite{decoherence}, and quantum technology \cite{Ladd,Gisin}. Many tools have been proposed for investigating the extent to which quantum mechanics can describe the rationale behind observed phenomena and ensure that key procedures are reliably performed in the quantum regime. Among these tools, entanglement~\cite{entsep}, Bell nonlocality \cite{Bellnon}, and {quantum discord \cite{Zurek,Vedral,review}} are the most commonly used. These measures adopt different standpoints to evaluate the quantum characteristics. For example, the quantum discord is based on the concept of mutual information. Notably, the quantum discord is capable of capturing quantum correlations not only in entangled states, but also in separable states for bipartite systems.

The main principle of quantum discord is to characterize and evaluate the difference between two expressions of the mutual information in bipartite systems. However, various approaches have also been proposed for evaluating quantum discord in systems involving more than two parties. {For~example, the study uses the monogamy relation \cite{Koashi} to expresses entanglement of formation in terms of different discord in multipartite systems \cite{Fanchini}.} Another study based on the relative entropy \cite{GQD} allows the measure of multipartite quantum discord to be nonnegative for arbitrary states. In addition, the quantum discord has been directly generalized by using multivariate mutual information \cite{dissension} to reveal various correlation features under the mutual information described in different ways. In the present study, tripartite quantum correlations are characterized and certified using a tripartite information diagram conditioned on one circle, where each circle represents a~different variable. Using the proposed measure, tripartite correlations are demonstrated for states close to Greenberger--Horne--Zeilinger, W, and biseparable states. Furthermore, a comparison is made between the multipartite discord correlation considered in this study and two other important quantum correlations---namely genuine multipartite entanglement \cite{nonsep} and genuine multipartite Einstein--Podolsky--Rosen steerability  \cite{steer}.

\section{Quantum Discord}

In classical information theory \cite{Infor}, the Shannon entropy for an unknown random variable $X$ is given as $H(X)=-\sum_{i}P(x_{i})\log_2{P(x_{i}})$, where $P(x_{i})$ is the probability of variable $x_i$. Moreover, the~correlation between two random variables $X$ and $Y$ is measured by their mutual information; i.e.,
\begin{equation}
J(X;Y)=H(X)-H(X|Y),
\end{equation}
where $H(X|Y)\equiv\sum_{j}P(y_{j})H(X|Y_{=y_{j}})$ is the conditional entropy of $X$ conditioned on the value $y_{j}$ of $Y$. Applying the Bayes rule \cite{Bayes}, the conditional entropy $H(X|Y)$ can be rewritten in terms of the entropy $H(X)$ and joint entropy $H(X,Y)=-\sum_{i}\sum_{j}P(x_i,y_j)\log_2{P(x_i,y_j)}$, where $P(x_i,y_j)$ is the joint probability of variables $x_i$ and $y_j$. In other words, the conditional entropy $H(X|Y)$ is equal to {$H(X,Y)-H(Y)$}. Consequently, the following classically equivalent expression for the mutual information can be obtained:
\begin{equation}
I(X;Y)=H(X)+H(Y)-H(X,Y).
\end{equation}
Thus, the mutual information $I(X;Y)$ and $J(X;Y)$ are equal in classical systems.

In the quantum regime, a composite system $AB$ consisting of subsystems $A$ and $B$ is described by the density operator $\rho_{AB}$. Furthermore, when tracing out a subsystem, the reduced density operators for subsystems $A$ and $B$ are represented by $\rho_A$ and $\rho_B$, respectively. The von Neumann entropy for the system expressed in the form of the density operator is $S(\rho)=-Tr[\rho\log_2{\rho}]$. For instance, the~entropy of subsystem $A$ is $S(A)=S(\rho_A)$, while that of the composite system $AB$ is $S(A,B)=S(\rho_{AB})$. Moreover, the conditional entropy of $A$ conditioned on the states of $B$ is $S(A|B)=\sum_jP_jS(\rho_{A|\Pi^j_B})$, where $\Pi^j_B$ is the $j$ projector set for subsystem $B$ and $P_j$ is the probability of measuring state $j$. The~mutual informations $I(A;B)$ and $J(A;B)$ are then given as \cite{Zurek}:
\begin{equation}\label{mutual}
\begin{split}
I(A;B)&=S(A)+S(B)-S(A,B),\\
J(A;B)&=S(A)-S(A|B).
\end{split}
\end{equation}

As shown in Equation~(\ref{mutual}), the conditional entropy $S(A|B)$ is not equal to the entropy \mbox{$S(A,B)-S(B)$} for quantum systems in general. The difference between the two expressions for the mutual information, $I(A;B)$ and $J(A;B)$, is referred to as the quantum discord \cite{Zurek,Vedral}, and is defined~as
\begin{equation}
\delta(A;B)=\min_{\Pi_B^j}[I(A;B)-J(A;B)].
\end{equation}

\begin{figure}
\includegraphics[width=8.5cm]{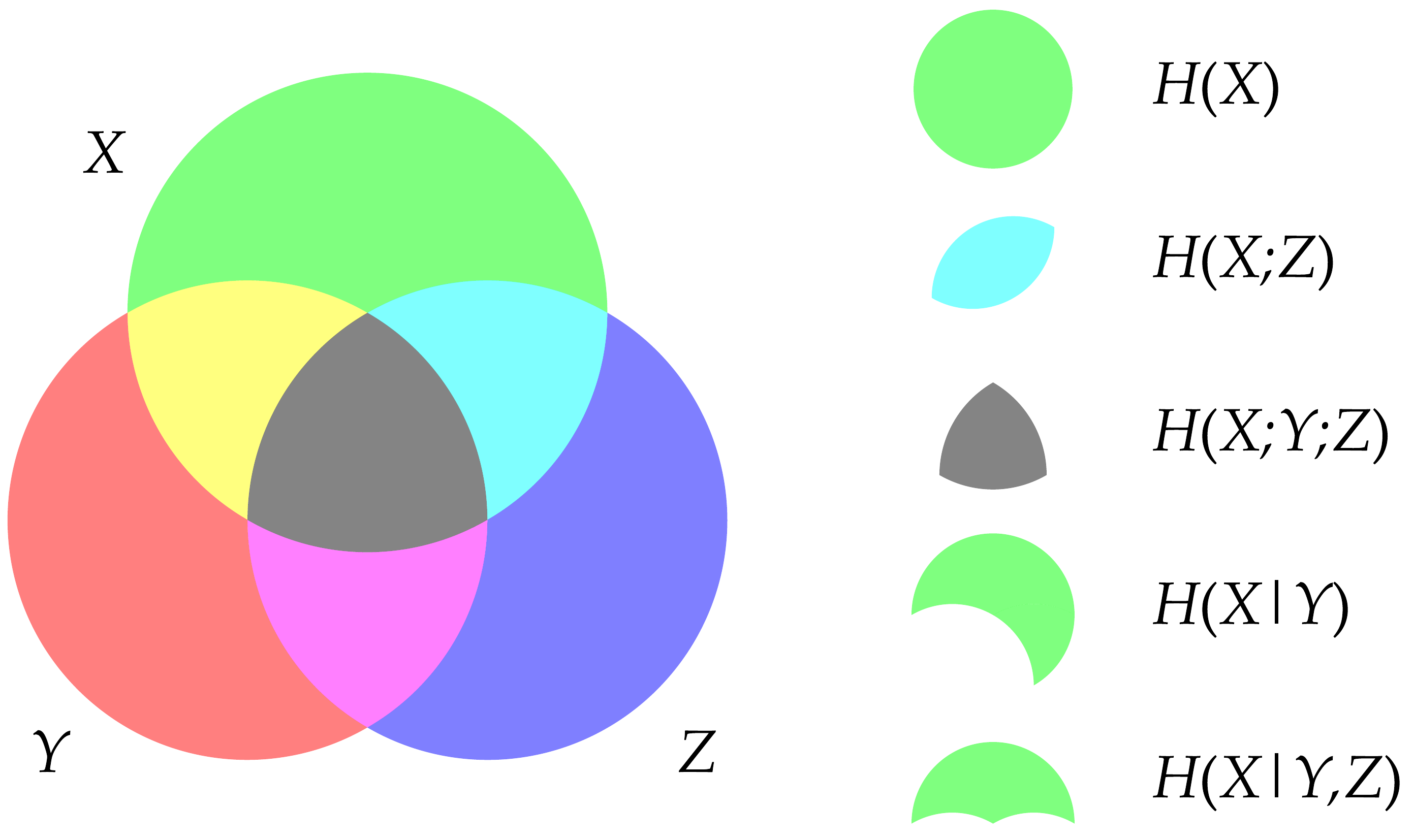}
\caption{Tripartite information diagram (TID). The three circles represent the individual entropies of  variables $X$, $Y$, and $Z$. The mutual information among the variables is represented by the overlapping region of the circles. The conditional entropies are represented by the circles of the corresponding variable excluding the overlaps of the circles corresponding to the conditioned variables. For instance, the mutual information of variables $X$ and $Z$ is represented by the cyan block, which is the overlap between the green circle and the blue circle. Similarly, the conditional entropy of $X$ conditioned on $Y$ and $Z$ is the green block excluding the yellow block and the cyan block.}
\label{TID}
\end{figure}

The value of function $I(A;B)-J(A;B)$ may increase due to projective measurements on $B$. This~increase causes an uncorrelated system to show a false signal of quantum discord. In order to eliminate such measurement-induced disturbance (MID) \cite{MID}, it is necessary to find the  projector $\Pi_B^j$ which minimizes the discord. The quantum discord \cite{Zurek,Vedral} can be applied to general bipartite systems, such as multi-qubit systems partitioned into two groups. However, it does not allow the correlations between each qubit to be described very clearly. Accordingly, the following section presents an~approach for certifying and characterizing the quantum discord in multipartite systems by means of a tripartite information diagram (TID).
\section{Revealing Tripartite Quantum Discord with Tripartite Information Diagram}

Consider the TID shown in Figure \ref{TID}, in which the three circles represent the entropies of three different variables, $X$, $Y$, and $Z$. As shown, the color block corresponding to the multivariate mutual information $H(X;Y;Z)$ can be described by more than two sets of combinations of the other color blocks. Consequently, more than two expressions for the multivariate mutual information exist; i.e.,
\begin{widetext}
\begin{equation}\label{Cmutual}
\begin{split}
H(X;Y;Z)&=H(X)+H(Y)+H(Z)-H(X,Y)-H(Y,Z)-H(X,Z)+H (X,Y,Z)\\
&=H(X,Y)-H(Y|X)-H(X|Y)-H(X|Z)-H(Y|Z)+H(X,Y|Z)\\
&=H(X)+H(Y)+H(Z)-H(X,Y)-H(X,Z)+H(X|Y,Z).
\end{split}
\end{equation}
\end{widetext}

It is noted that all of these expressions are equal in classical systems.

In the present study, the tripartite quantum discord is defined as the difference between two expressions of the mutual information derived from the TID. Furthermore, the variables $X$, $Y$, and~$Z$ are treated as subsystems $A$, $B$, and $C$, and~their entropies are described as von Neumann entropy. Since many different expressions for classical mutual information exist, tripartite quantum discord can also be described in many formats. For instance, quantum dissension \cite{dissension} can be described by the quantum analogue of Equation~(\ref{Cmutual}). This study presents a further format for describing tripartite quantum discord. Specifically, one of the expressions of mutual information $I(A;B;C)$ is defined as comprising entropies of reduced systems and joint entropies only; that is,
\begin{eqnarray}
&&I(A;B;C)=S(A)+S(B)+S(C)-S(A,B)\nonumber\\
&&\ \ \ \ \ \ \ \ \ \ \ \ \ \ \ \ \ \ \ -S(B,C)-S(A,C)+S(A,B,C).
\end{eqnarray}
Referring again to Figure \ref{TID}, the aim for the other expression of mutual information is to construct the multivariate mutual information without joint entropies in one circle, where each circle represents the entropy of one variable. For example, $H(X;Y;Z)$ consists of $H(X)$, $H(X|Y)$, $H(X|Z)$, and $H(X|Y,Z)$. Given a system with three variables, there exist three expressions for the multivariate mutual information. Let the three expressions be defined as the quantum analogues $J_k(A;B;C)$ with $k\in\{A,B,C\}$ shown as below:
\begin{widetext}
\begin{equation}\label{7}
\begin{split}
&J_A (A;B;C)=S(A)-S(A|B)-S(A|C)+S(A|B,C),\\
&J_B (A;B;C)=S(B)-S(B|A)-S(B|C)+S(B|A,C),\\
&J_C (A;B;C)=S(C)-S(C|B)-S(C|A)+S(C|A,B).
\end{split}
\end{equation}
\end{widetext}
The mutual information $J_k(A;B;C)$ comprises entropies of reduced systems and conditional entropies including one-particle projective measurements and two-particle projective measurements. For~each expression $J_k(A;B;C)$, one specific subsystem is not projected. However, those projective measurements which are performed may cause MID. Thus, to eliminate potential false signals of tripartite quantum discord, a search is made for the projectors $\Pi_m$ and $\Pi_n$, where $m,n \in \{A,B,C\}$ and $m,n\neq k$, which minimize the function $I(A;B;C)-J_k(A;B;C)$. As a result, the proposed tripartite quantum discord measure is defined as
\begin{equation}\label{Dis}
\delta_k(A;B;C)=\min\limits_{\Pi_m,\Pi_n}[I(A;B;C)-J_k(A;B;C)].
\end{equation}
{For the projectors $\Pi_m$ and $\Pi_n$, in the present work we concern how  discord can be observed by performing local measurements on spatially separated subsystems. Such scenarios can have direct connections with the quantum-information tasks such as quantum communication \cite{Gisin} and one-way quantum computation \cite{Ladd}.} 

\begin{figure}
\includegraphics[width=7.5cm]{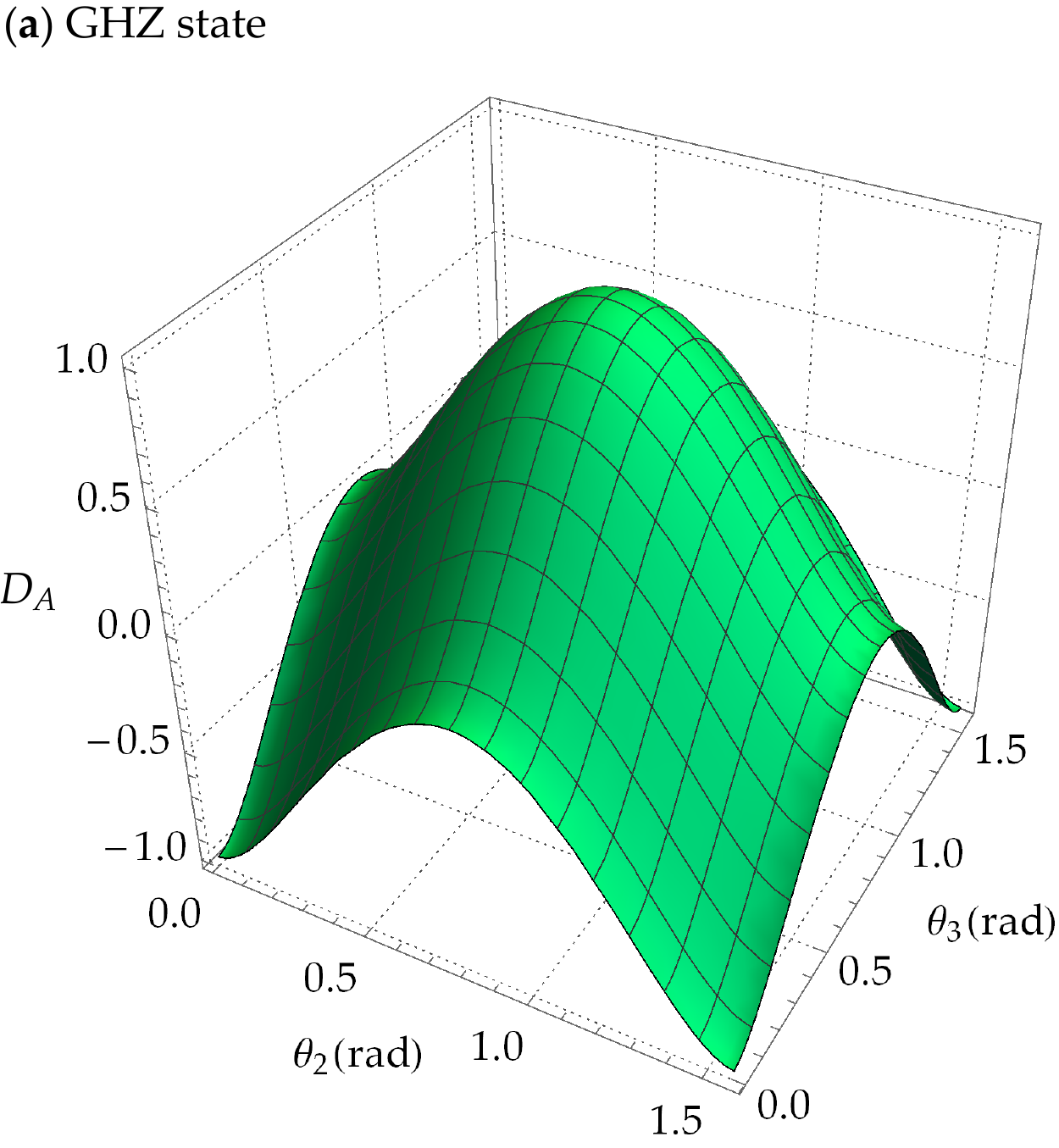}
\includegraphics[width=7.5cm]{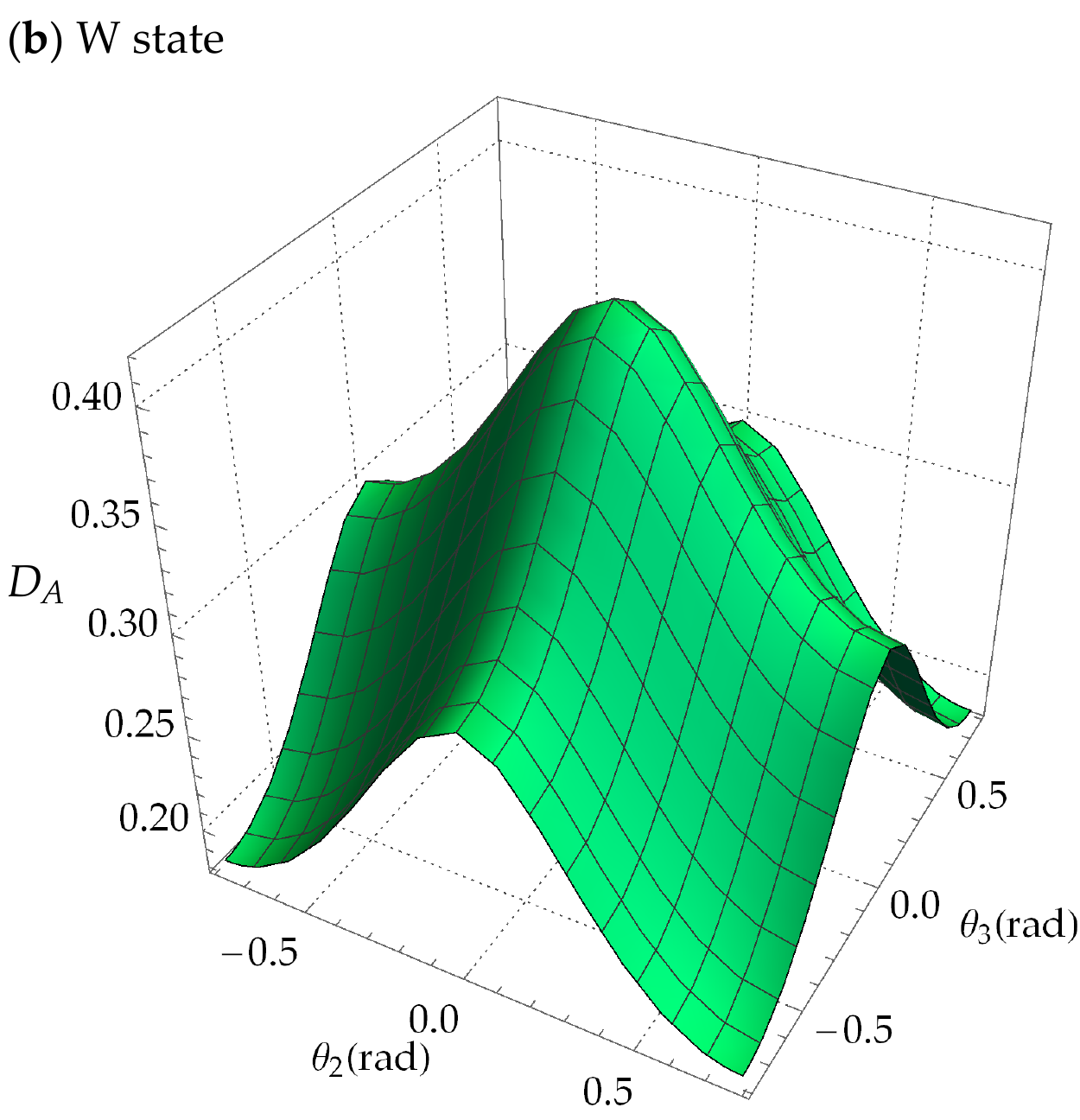}
\caption{Tripartite quantum discord focused on one subsystem for pure states. (\textbf{a}) For the Greenberger-- Horne--Zeilinger (GHZ) state, the minimum of function $D_A(\theta_2,\theta_3)$ occurs at $\theta_2=\theta_3=0$; (\textbf{b}) for the W state, the minimum of function $D_A(\theta_2,\theta_3)$ occurs at $\theta_2=\theta_3=\pi/4$.}
\label{PureState}
\end{figure}

{It is worth noting that, compared to the discord for bipartite systems \cite{Zurek}, $\delta_k(A;B;C)$ can be negative. The reason is that measurements on subsystems for conditional entropies can make a~pure state of the multipartite system collapse to a mixed state, and can increase the uncertainty of the remaining~subsystems of interest.~This typically involves an increase of entropy.~Hence, we have \mbox{$S(\alpha|\beta)\geq S(\alpha,\beta)-S(\beta)$} and $S(\alpha|\beta,\gamma)\geq S(\alpha,\beta,\gamma)-S(\beta,\gamma)$ for $\alpha,\beta,\gamma\in\{A,B,C\}$ and \mbox{$\alpha\neq\beta\neq\gamma$ \cite{Zurek}}. Both inequalities have effects on the function $\delta_k(A;B;C)$. As a result, the value of tripartite quantum discord can be negative for quantum systems. See Section \ref{sec5} and Figure~\ref{PureState} for concrete examples. Furthermore, one may consider that the discord can be alternatively defined as: $\delta(A;B;C)=\min_k\delta_k(A;B;C)$, to describe possible integral correlation of a tripartite system.~However, for biseparable systems as will be illustrated in Section \ref{sec5.4}, the tripartite quantum discord $\delta(A;B;C)$ derived from this definition cannot be detected when the subsystem $k$ is separate from the other two subsystems, whereas our measure can still show the discord of biseparable systems.}
Referring to the framework of the proposed measure, the mutual information $I(A;B;C)$ obeys the inclusion--exclusion principle. Furthermore, the mutual information $J(A;B;C)$ can be described by rewriting the mutual information $I(A;B;C)$ with the concept of Bayes rule \cite{Bayes} used in classical systems. Therefore, the proposed measure can be extended to general N-partite systems.

\section{Tripartite Quantum Discord Focused on One Subsystem for Pure States}

In accordance with Equation~(\ref{7}), focusing on one subsystem (e.g., $A$) means that the mutual information $J_A(A;B;C)$ only comprises entropy of subsystem $A$ and conditional entropies of subsystem~$A$ conditioned on the other subsystems. In order to investigate the discord in tripartite quantum systems, the following discussions consider two particular pure tripartite states; namely, the~Greenberger--Horne--Zeilinger (GHZ) state \cite{GHZ} and the W state.

\subsection{Tripartite Quantum Discord Focused on One Subsystem for GHZ State}
Consider a pure tripartite GHZ state \cite{GHZ} of the form
\begin{equation}
\left|GHZ\right\rangle=\frac{1}{\sqrt{2}}(\left|0_A0_B0_C\right\rangle+\left|1_A1_B1_C\right\rangle),
\end{equation}
where $\left|0\right\rangle$ and $\left|1\right\rangle$ are computational basis states and form an orthonormal basis for the vector space. After tracing out two subsystems, the density operators representing the individual subsystems are given by
\begin{equation}\label{10}
\rho_A=\rho_B=\rho_C=\frac{1}{2}(\left|0\right\rangle\!\!\left\langle0\right|+\left|1\right\rangle\!\!\left\langle1\right|),
\end{equation}
where the von Neumann entropies are all equal to one; i.e., $S(A)=S(B)=S(C)=1$. Similarly,~by~tracing out any one of the three subsystems, the reduced density operators are obtained as
\begin{equation}\label{11}
\rho_{AB}=\rho_{BC}=\rho_{AC}=\frac{1}{2}(\left|00\right\rangle\!\!\left\langle00\right|+\left|11\right\rangle\!\!\left\langle11\right|).
\end{equation}
The entropies of the three density operators are given by $S(A,B)=S(B,C)=S(A,C)=1$, respectively. Since the GHZ state is a pure state, the joint von Neumann entropy $S(A,B,C)$ is equal to~zero.

To define the local measurements, consider the following rotations in the directions of the basis vectors of subsystems:
\begin{eqnarray}
\left|+\right\rangle_j&=&\text{cos}(\theta_j)\left|0\right\rangle_j+e^{i\phi_j}\text{sin}(\theta_j)\left|1\right\rangle_j,\\
\left|-\right\rangle_j&=&\text{cos}(\theta_j)\left|1\right\rangle_j-e ^{i\phi_j}\text{sin}(\theta_j)\left|0\right\rangle_j,
\end{eqnarray}
where $j=1,2$, and $3$ for subsystems $A$, $B$, and $C$, respectively. The conditional entropies of one subsystem conditioned on another subsystem for the GHZ state can then be represented by
\begin{eqnarray}\label{14}
&&S(\alpha|\beta)=-(\frac{1+\text{cos}(2\theta_j)}{2})\log_2{\frac{1+\text{cos}(2\theta_j)}{2}}\nonumber\\
&& \ \ \ \ \ \ \ \ \ \ \ \ \ \ -(\frac{1-\text{cos}(2\theta_j)}{2})\log_2{\frac{1-\text{cos}(2\theta_j)}{2}},\quad \alpha\neq\beta,\nonumber\\
\end{eqnarray}
for $\alpha,\beta \in \{A,B,C\}$. Furthermore, the conditional entropies of one subsystem given the other two subsystems is reduced to zero. Referring to Equation~(\ref{14}), and taking $\delta_A(A;B;C)$ for illustration purposes, the tripartite quantum discord $\delta_A(A;B;C)$ can be obtained by minimizing function $D_A(\theta_2,\theta_3)$ over the angles $\theta_2$ and $\theta_3$. The function  to be minimized is given by
\begin{widetext}
\begin{equation}
\begin{aligned}
D_A(\theta_2,\theta_3)=[-1-&(\frac{1+\text{cos}(2\theta_2)}{2})\log_2{\frac{1+\text{cos}(2\theta_2)}{2}}-(\frac{1-\text{cos}(2\theta_2)}{2})\log_2{\frac{1-\text{cos}(2\theta_2)}{2}}\\
-&(\frac{1+\text{cos}(2\theta_3)}{2})\log_2{\frac{1+\text{cos}(2\theta_3)}{2}}-(\frac{1-\text{cos}(2\theta_3)}{2})\log_2{\frac{1-\text{cos}(2\theta_3)}{2}}].
\end{aligned}
\end{equation}
\end{widetext}

As seen in Figure \ref{PureState}a, by minimizing over the angles $\theta_2$ and $\theta_3$, the tripartite quantum discord is obtained as $\delta_A(A;B;C)=-1$. In other words, unlike the original quantum discord, the tripartite quantum discord measure can have a negative value.

\subsection{Tripartite Quantum Discord Focused on One Subsystem for W State}

The state vector of W state has the form
\begin{equation}\label{16}
\left|W\right\rangle=\frac{1}{\sqrt{3}}(\left|1_A0_B0_C\right\rangle+\left|0_A1_B0_C\right\rangle+\left|0_A0_B1_C\right\rangle).
\end{equation}
By tracing out two subsystems, the individual subsystems are obtained as
\begin{equation}\label{17}
\rho_{A}=\rho_{B}=\rho_{C}=\frac{1}{3}(2\left|0\right\rangle\!\!\left\langle0\right|+\left|1\right\rangle\!\!\left\langle1\right|).
\end{equation}
The von Neumann entropies of three subsystems are all equal to $0.918$. By tracing out any one of the three subsystems, the reduced density operators of the W state are obtained as
\begin{eqnarray}\label{18}
&&\rho_{AB}=\rho_{BC}=\rho_{AC}\nonumber\\
&&=\frac{1}{3}(\left|00\right\rangle\!\!\left\langle00\right|+\left|01\right\rangle\!\!\left\langle01\right|+\left|01\right\rangle\!\!\left\langle10\right|+\left|10\right\rangle\!\!\left\langle01\right|+\left|10\right\rangle\!\!\left\langle10\right|).\nonumber\\
\end{eqnarray}
Hence, the von Neumann entropies of the reduced two subsystems, $S(A,B$), $S(B,C)$, and $S(A,C)$, are~$0.918$ and the joint entropy $S(A,B,C)$ is equal to zero for the purity of W state (Equation (\ref{16})).

The conditional entropies based on local measurements for the W state are given as
\begin{equation}\label{19}
\begin{aligned}
S(\alpha|\beta)=\sum_jP_jS(\rho_{\alpha|\Pi_\beta^j})&=-\sum_j(\lambda_j\log_2{\lambda_j)}, \quad \alpha\neq\beta,
\end{aligned}
\end{equation}
for $\alpha,\beta \in \{A,B,C\}$, where $\lambda_j$ are the eigenvalues of subsystem $\alpha$ after subsystem $\beta$ is projected with the $j$ projector and $P_j$ is the probability of measuring state $j$. The conditional entropy of one subsystem conditioned on the other two subsystems is obviously reduced to zero, Referring to Equations~(\ref{17})--(\ref{19}), and taking $\delta_A (A;B;C)$ for illustration purposes, the function $I(A;B;C)-J_A(A;B;C)$ can be described by the function $D_A(\theta_2,\theta_3)$.~The tripartite quantum discord $\delta_A (A;B;C)$ is then obtained via the minimization of function $D_A(\theta_2,\theta_3)$.

As shown in Figure \ref{PureState}b, the tripartite quantum discord focused on subsystem $A$ is equal to $0.182$. The interference of the MID is similar to that of the pure tripartite GHZ state. However, for the pure W state, the tripartite quantum discord is nonnegative for arbitrary projectors.

\section{Tripartite Quantum Discord Focused on One Subsystem for the Mixed States}   \label{sec5}

In reality, the states of a system are not pure, but contain noise {induced} by environment. Accordingly, this section takes the mixed states in the form of {Werner state \cite{werner}} to evaluate the effect of the state purity on the performance of the proposed tripartite quantum discord measure. In addition, the connection between tripartite quantum discord and two other types of quantum correlation---namely genuine tripartite entanglement (GTE) and genuine tripartite Einstein--Podolsky--Rosen steering (GTEPRS)---are briefly discussed.

\subsection{Tripartite Quantum Discord Focused on One Subsystem for the Werner-GHZ States}

Consider the following Werner-GHZ state:
\begin{equation}
\rho_{GHZ}=\frac{1-\mu}{8}\mathcal{I}+\mu\left|GHZ\right\rangle\!\!\left\langle GHZ\right|,
\end{equation}
where $\mu$ is the purity of the Werner-GHZ state $(0\leq\mu\leq1)$ and $\mathcal{I}$ is the identity operator. As in the case of pure tripartite GHZ state, $S(A)=S(B)=S(C)=1$, since $\rho_A$, $\rho_B$, and $\rho_C$ are all half of identity operator. By tracing out a single subsystem, the reduced subsystems are obtained as
\begin{eqnarray}
&&\rho_{AB}=\rho_{BC}=\rho_{AC}\nonumber\\
&&=\frac{1+\mu}{4}(\left|00\right\rangle\!\!\left\langle 00\right|+\left|11\right\rangle\!\!\left\langle11\right|)+\frac{1+\mu}{4}(\left|01\right\rangle\!\!\left\langle01\right|+\left|10\right\rangle\!\!\left\langle10\right|).\nonumber\\
\end{eqnarray}

Let local measurements be taken in the bases which minimize the function $D_A(\theta_2,\theta_3)$ for the pure tripartite GHZ state. The conditional entropies of one subsystem conditioned on another subsystem for the Werner-GHZ state can then be represented as
\begin{eqnarray}\label{OneCon}
&&S(A|B)=S(A|C)=-(\frac{1+\mu}{2})\log_2{(\frac{1+\mu}{2})}\nonumber\\
&&\ \ \ \ \ \ \ \ \ \ \ \ \ \ -(\frac{1-\mu}{2})\log_2{(\frac{1-\mu}{2})}.
\end{eqnarray}
Similarly, the conditional entropy of one subsystem conditioned on the other two subsystems is obtained as
\begin{eqnarray}\label{TwoCon}
&&S(A|B,C)=\frac{1-\mu}{2}-(\frac{1-\mu}{4})\log_2{(\frac{1-\mu}{2(1+\mu)})}\nonumber\\
&&\ \ \ \ \ \ \ \ \ \ \ \ \ \ \ \ \ \ -(\frac{1+3\mu}{4})\log_2{(\frac{1+3\mu}{2(1+\mu)})}.
\end{eqnarray}
The tripartite quantum discord $\delta_A(A;B;C)$ can then be obtained by substituting Equations (\ref{OneCon}) and~(\ref{TwoCon}) into Equation~(\ref{Dis}).

Figure \ref{Wernerstate}a plots $\delta_A(A;B;C)$ as a function of the state purity $\mu$. As shown, the tripartite quantum discord approaches zero as the state purity reduces, and vanishes at $\mu=0$ (i.e., $\rho_{GHZ}$ is a completely mixed state).

\subsection{Tripartite Quantum Discord Focused on One Subsystem for the Werner-W States}

The Werner-W state has the form
\begin{equation}
\rho_{W}=\frac{1-\mu}{8}\mathcal{I}+\mu\left|W\right\rangle\!\!\left\langle W\right|,
\end{equation}
where $\mu$ is the purity of the W state ($0\leq\mu\leq1$). The reduced density operators of the single subsystems are given by
\begin{equation}
\rho_A=\rho_B=\rho_C=\frac{3+\mu}{6}\left|0\right\rangle\!\!\left\langle0\right|+\frac{3-\mu}{6}\left|1\right\rangle\!\!\left\langle1\right|\,.
\end{equation}
Similarly, the density operators of the two subsystems obtained by tracing out a single subsystem are given as
\begin{equation}
\begin{aligned}
\rho_{AB}=&\rho_{BC}=\rho_{AC}=\frac{3+\mu}{12}(\left|00\right\rangle\!\!\left\langle00\right|+\left|01\right\rangle\!\!\left\langle01\right|+\left|10\right\rangle\!\!\left\langle10\right|)\\
+&\frac{1-\mu}{4}\left|11\right\rangle\!\!\left\langle11\right|+\frac{\mu}{3}(\left|01\right\rangle\!\!\left\langle10\right|+\left|10\right\rangle\!\!\left\langle01\right|).
\end{aligned}
\end{equation}

The set of local measurement bases originates from the tripartite quantum discord for the pure W~state. The conditional entropies of one subsystem conditioned on another subsystem for the Werner-W state are expressed as
\begin{eqnarray}\label{24}
&&S(A|B)=S(A|C)=-(\frac{3-\sqrt{5}\mu}{6})\log_2{(\frac{3-\sqrt{5}\mu}{6})}\nonumber\\
&&\ \ \ \ \ \ \ \ \ \ \ \ \ -(\frac{3+\sqrt{5}\mu}{6})\log_2{(\frac{3+\sqrt{5}\mu}{6})}.
\end{eqnarray}
Similarly, the conditional entropy of one subsystem conditioned on the other two subsystems is obtained as
\begin{widetext}
\begin{equation}\label{25}
\begin{aligned}
S(A|B,C)=-&(\frac{1-\mu}{4})\log_2{(\frac{3(1-\mu)}{3+2\mu})}
-(\frac{3+7\mu)}{12})\log_2{(\frac{3+7\mu)}{2(3+2\mu)})}\\
-&(\frac{1-\mu}{4})\log_2{(\frac{3(1-\mu)}{2(3-2\mu)})}
-(\frac{3-\mu}{12})\log_2{(\frac{3-\mu}{2(3-2\mu)})}.
\end{aligned}
\end{equation}
\end{widetext}
Inserting Equations~(\ref{24}) and (\ref{25}) into Equation~(\ref{Dis}), we can find the value of $\delta_A(A:B:C)$.

\begin{figure}
\centering
\includegraphics[width=7.5 cm]{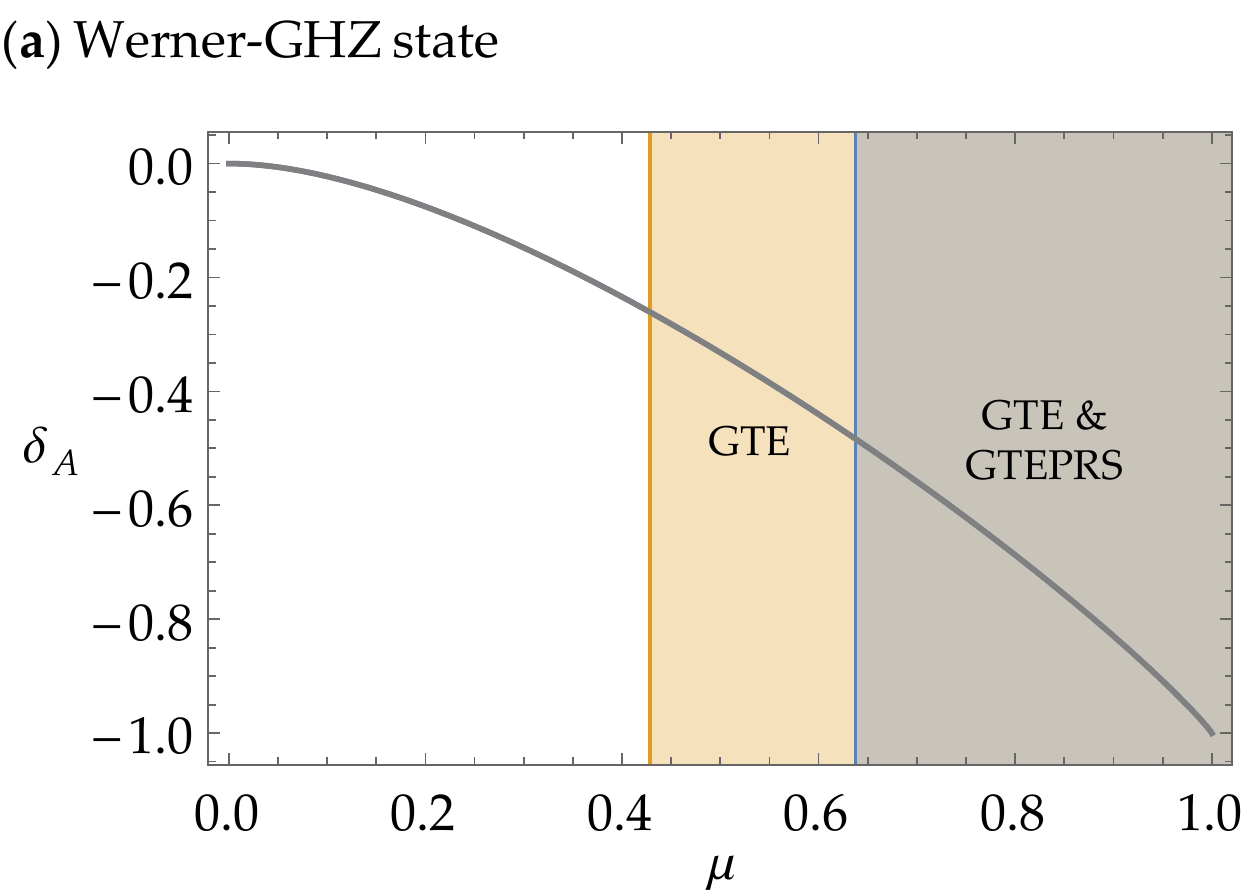}
\includegraphics[width=7.5 cm]{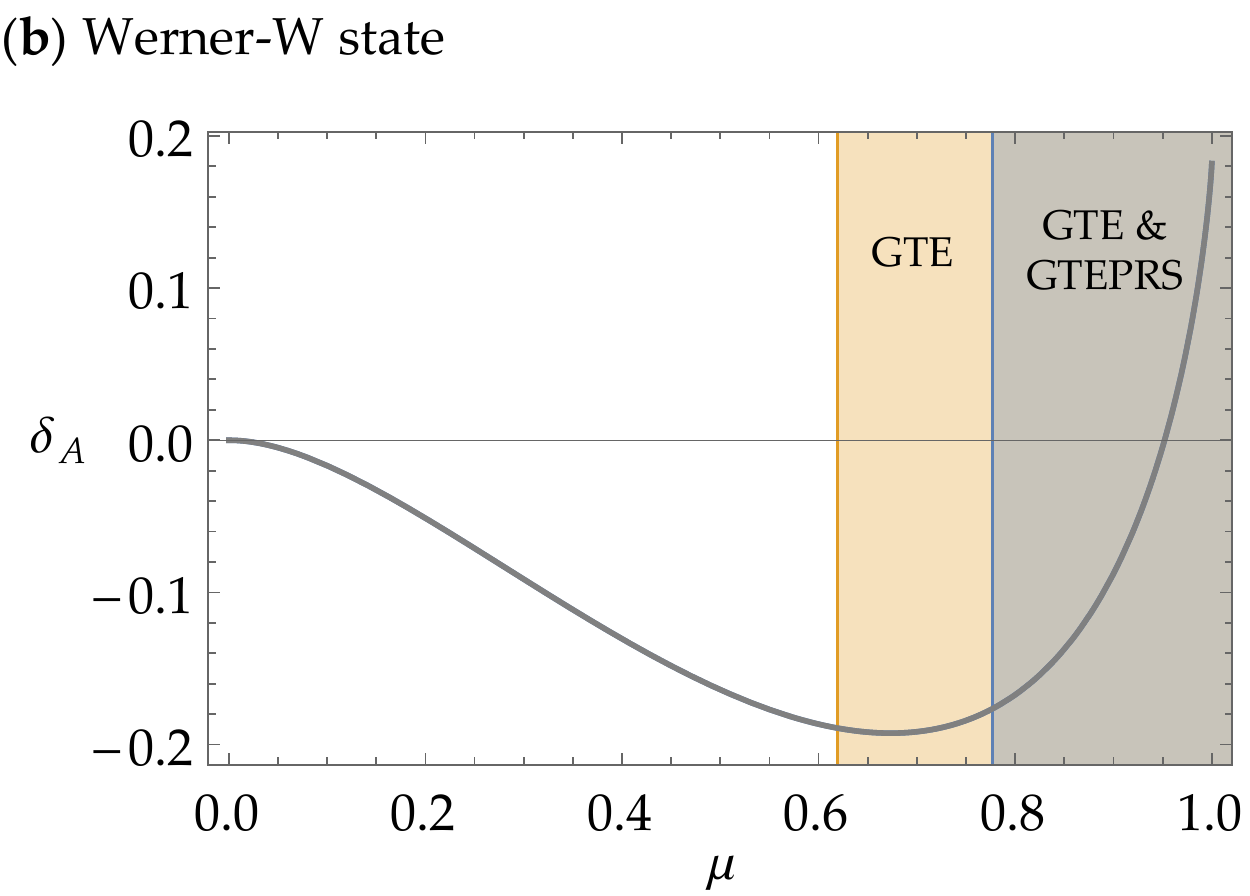}
\caption{ Tripartite quantum discord focused on one subsystem for Werner states. (\textbf{a}) $\delta_A (A;B;C)$ for Werner-GHZ state as function of $\mu$; (\textbf{b}) $\delta_A (A;B;C)$ for Werner-W state as function of $\mu$. With the boundaries derived from \cite{nonsep,steer}, the orange line shows the boundary of genuine tripartite entanglement (GTE) while the blue line shows the boundary of genuine tripartite Einstein--Podolsky--Rosen steering~(GTEPRS).}
\label{Wernerstate}
\end{figure}

Figure \ref{Wernerstate}b plots $\delta_A(A;B;C)$ as a function of $\mu$. As shown, the tripartite quantum discord is equal to zero not only in the completely mixed state, but also in a close-to-pure state. The reason is that the value of $S(A|B)-S(A,B)+S(B)+S(A|C)-S(A,C)+S(C)$ is equal to that of \mbox{$S(A|B,C)-S(A,B,C)+S(B,C)$}.

\subsection{Relation with Genuine Tripartite Entanglement and Einstein--Podolsky--Rosen (EPR) Steering}

This section briefly reviews the genuine multipartite entanglement (GME) \cite{nonsep} and genuine multipartite Einstein--Podolsky--Rosen steering (GMEPRS) \cite{steer}, and discusses their applications in tripartite quantum systems for the purpose of comparison with the proposed tripartite quantum~discord.

For GME, a witness operator $\mathcal{W}$ that detects the GME for states close to $\left|\psi\right\rangle$ is given by \cite{nonsep}
\begin{equation}\label{GME}
\mathcal{W}_{GME}\equiv\alpha_{GME}\mathcal{I}-\left|\psi\right\rangle\!\left\langle\psi\right|,
\end{equation}
where $\alpha_{GME}$ is the maximum overlap between the biseparable state and the pure state $\left|\psi\right\rangle$. Given the witness operator, all biseparable states $\rho_{bi}$ satisfy $Tr(\mathcal{W}_{GME}\rho_{bi})\geq0$. If a state $\rho$ that $Tr(\mathcal{W}_{GME}\rho)<0$, the state is identified as GTE.

In tripartite system, the maximum overlap $\alpha_{GME}$ for GHZ state and W state are $1/2$ and $2/3$, respectively. Referring to Figure \ref{Wernerstate}, and representing the witness operator by the purity $\mu$, the~boundary for Werner-GHZ state is $3/7$ while that for the Werner-W state is $13/21$. It can further be shown that if the tripartite quantum discord $\delta_A(A;B;C)$ is less than $-0.261$ the Werner-GHZ state shows GTE. However, the tripartite quantum discord for the Werner-W state may have the same value on both sides of the boundary, in which case $\delta_A(A;B;C)$ cannot indicate the boundary of GME.

Regarding the GMEPRS, assume that a system contains $N$ parties and a source is capable of creating $N$-particles. Assume further that each party can receive a particle from the source whenever an $N$-particle state is created. Let the system be divided into two groups, $A_s$ and $B_s$, where $A_s$ is responsible for sending particles from the source to every party. After receiving particles, the parties measure their respective parts and communicate classically. Since $B_s$ does not trust $A_s$, $A_s$ need to convince $B_s$ that the state shared between them is entangled. $A_s$ performs this task if and only if it can prepare different ensembles of quantum states for $B_s$ by steering $B_s$’s state. The $N$-particle state is genuine $N$-partite EPR steering if $A_s$ can achieve the task for all bipartitions $A_s$ and $B_s$ of the $N$-particle system. Given full information about a target state $\left|\psi\right\rangle$, the witness has the form \cite{steer}:
\begin{equation}\label{GMEPRS}
\mathcal{W}_{EPR}\equiv\alpha_{EPR}\mathcal{I}-\left|\psi\right\rangle\!\left\langle\psi\right|,
\end{equation}
with critical witness kernel $\alpha_{EPR}\equiv\max_{v_1^{(m_1)},..., v_N^{(m_N)}}$ ${\sum_{v_1^{(m_1)}...v_N^{(m_N)}}}c(v_1^{(m_1)}...v_N^{(m_N)})P(v_1^{(m_1)},..., v_N^{(m_N)})$, where $c(v_1^{(m_1)}...v_N^{(m_N)})$ are derived from the tomographic decomposition of state $\left|\psi\right\rangle$ and the joint probability $P(v_1^{(m_1)}, \dots, v_N^{(m_N)})$ satisfies that one group has a preexisting-state scenario while the other group performs quantum measurements on preexisting quantum states. If a state $\rho$ that $Tr(\mathcal{W}_{EPR}\rho)<0$, the state is identified as GMEPRS.

In tripartite systems, the critical witness kernel $\alpha_{EPR}$ for GHZ state and W state are $0.683$ and $0.8047$, respectively. Let the GMEPRS witness be represented by the purity parameter $\mu$ in Figure \ref{Wernerstate}. It is observed that if the tripartite quantum discord $\delta_A(A;B;C)=-0.484$ and the purity is greater than $0.6377$, the Werner-GHZ state shows GTEPRS. For the Werner-W state, the boundary lies at $\mu=0.7768$. Since $\delta_A(A;B;C)$ can have the same value on both sides of the boundary, the tripartite quantum discord cannot definitely identify GTEPRS based on the state purity. However, a state $\rho$ can absolutely show GTEPRS when $\delta_A(A;B;C)$ is greater than zero.

\subsection{Tripartite Quantum Discord Focused on One Subsystem for Biseparable States}   \label{sec5.4}

Unlike Werner-GHZ states or Werner-W states, the structure of biseparable states may be asymmetric for subsystems $A$, $B$, and $C$. Thus, the tripartite quantum discord $\delta_k(A;B;C)$ for a~biseparable state may be different from that of subsystem focused on. For example, a biseparable state is expressed as
\begin{equation}
\rho_{bi}=a\left|0\right\rangle_A\!\left\langle0\right|\otimes\left|\varphi^{+}\right\rangle_{BC}\!\left\langle\varphi^{+}\right|+b\left|1\right\rangle_A\!\left\langle1\right|\otimes\left|\psi^{-}\right\rangle_{BC}\!\left\langle\psi^{-}\right|,
\end{equation}
where $a+b=1$, $\left|\varphi^{+}\right\rangle=(\left|00\right\rangle+\left|11\right\rangle)/\sqrt{2}$, and $\left|\psi^{-}\right\rangle=(\left|01\right\rangle-\left|10\right\rangle)/\sqrt{2}$. It is easily shown that $\delta_A(A;B;C)$ is equal to zero, but $\delta_B(A;B;C)$ and $\delta_C(A;B;C)$ do not vanish. Indeed, due to its focus on just one subsystem, the proposed measure is able to distinguish which particular subsystem is separate from the others.

\section{Comparison with Other Measures}

This section briefly compares the multipartite quantum discord proposed in this study with two~other measures, namely global quantum discord \cite{GQD} and quantum dissension \cite{dissension}. The former is based on the relative entropy, and is a symmetric measure with a nonnegative value for an arbitrary state. By contrast, the quantum dissension and the present measure are based on the original mutual information concept proposed in \cite{Zurek} and are asymmetric measure. That is, both measures retain the characteristics of the original quantum discord for bipartite systems, in which projective measurements project on specific subsystems. However, the two measures are proposed for different purposes. In~particular, the quantum dissension is used to detect a system with a specific number of projective measurements, while that proposed in this study focus on one subsystem, and is thus suitable for the analysis of systems in which the projective measurements do not project on one specific subsystem.

\section{Conclusions}

This study has proposed a new measure for the identification of tripartite quantum correlations based on the tripartite information diagram and focused on just one subsystem. The feasibility of the proposed measure has been demonstrated for both Werner-GHZ states and Werner-W states. Moreover, the tripartite quantum correlation has been compared with the genuine multipartite nonseparability and steerability. Finally, it has been shown that the measure is capable of revealing the characteristics of biseparable states. Notably, the measure proposed in this study for revealing tripartite quantum discord can be readily extended to certify multipartite quantum discord focused on one subsystem by the Bayes rule \cite{Bayes} and the inclusion--exclusion principle.

\section{Acknowledgements}
This work is partially supported by the Ministry of Science and Technology, Taiwan, under Grant Numbers MOST 104-2112-M-006-016-MY3.

\end{document}